\documentclass{elsarticle}%
\usepackage{amssymb}
\usepackage{epsfig}
\usepackage{graphicx}
\usepackage{subfigure}
\usepackage{hyperref}
\usepackage{balance}
\biboptions{numbers,sort&compress}

\begin{document}
\unitlength 1 cm
\newtheorem{thm}{Theorem}
\newtheorem{lem}[thm]{Lemma}
\newdefinition{rmk}{Remark}
\newproof{pf}{Proof}
\newproof{pot}{Proof of Theorem \ref{thm2}}
\newcommand{\be}{\begin{equation}}
\newcommand{\ee}{\end{equation}}
\newcommand{\bearr}{\begin{eqnarray}}
\newcommand{\eearr}{\end{eqnarray}}
\newcommand{\nn}{\nonumber}
\newcommand{\vk}{\vec k}
\newcommand{\vp}{\vec p}
\newcommand{\vq}{\vec q}
\newcommand{\vkp}{\vec {k'}}
\newcommand{\vpp}{\vec {p'}}
\newcommand{\vqp}{\vec {q'}}
\newcommand{\bk}{{\bf k}}
\newcommand{\bp}{{\bf p}}
\newcommand{\bq}{{\bf q}}
\newcommand{\br}{{\bf r}}
\newcommand{\up}{\uparrow}
\newcommand{\down}{\downarrow}
\newcommand{\fns}{\footnotesize}
\newcommand{\ns}{\normalsize}
\newcommand{\cdag}{c^{\dagger}}

\title{Exact Diagonalization Study of 2D Hubbard Model on Honeycomb Lattice: Semi-metal-Insulator Transition}
\author{M. Ebrahimkhas}
\cortext[cor]{Corresponding author: tel.: +982182884489}
\ead{ebrahimkhas@gmail.com}
\address{Department of Science, Islamic Azad University, Mahabad Branch, Mahabad 59135, Iran}

\begin{abstract}

Phase transition in a honeycomb lattice is studied by the means of the two dimensional Hubbard model and the exact diagonalization dynamical mean field theory at zero temperature. At low energies,
the dispersion relation is shown to be a linear function of the momentum. In the limit of weak interactions, the system is in the
semi-metal phase. By increasing the on site interaction 
a semi-metal to insulator transition takes place in the paramagnetic phase. Calculation of double occupancy shows such a phase
transition is of the second order.  The respective phase transition point and critical on-site interaction are determined using
renormalized Fermi velocity factor.
\end{abstract}


\maketitle

\section{Introduction}
Materials with two dimensional honeycomb lattice represent an interesting class of low dimensional systems with many intriguing electronic structure properties. One such example is Graphene in which the conducting electrons act as massless fermions at low temperatures.  Owing to this unique feature~\cite{NetoRMP}, a number of remarkable phenomena such as the topological Mott-insulator transition \cite{Zhang-MIT} and 
quantum spin liquid \cite{Assad}, \cite{Baskaran-Jafari} have already been discovered to be realizable in this two dimensional honeycomb system. The dispersion relation corresponding to these  massless  Dirac fermions at low energies near the Fermi level can be described by a relativistic Hubbard-like Hamiltonian linear in terms of the momentum vector $k$  \cite{Semenoff}.\\

The interaction term in the Hubbard model plays a crucial role. 
In non interaction case, the honycomb  system is in a half-filling semi-metal phase. 
Increasing the on-site interaction in the system, many scenarios can be realized. For example, a number of methods  including, the Gutzwiller, quantum
Monte Carlo (QMC), iterative perturbation theory (IPT) and exact diagonalization (ED) methods at zero or finite temperature  
predict semi-metal to insulator transition (SMIT) or semi-metal to anti-ferromagnetic Mott insulator transition at different critical interactions\cite{Semenoff,Sorella,Santoro,Martello,Paiva,Jafari,Tran}. 
Among these methods, the so-called dynamical mean field theory (DMFT) is proven to be a powerful tool, allowing us  to find metal to insulator transition in the context of the strongly correlated systems. This method is exact at infinite dimension \cite{Georges-RMP}. In finite dimension, however, 
DMFT neglects nonlocal correlations. By exclusion of this term, one can, for example, notice a difference in the value of the on-site Coulomb term    $U_c$ as predicted by 
the  cluster dynamical mean field theory (CDMFT) and as obtained by DMFT   \cite{Wu}.\\
  
  In this work, the phase SMIT in honeycomb lattice is studied using DMFT coupled with ED method  at zero temperature. In this method, we first map Hubbard model to Anderson impurity model and create a single impurity model. The new model  is then solved exactly with an effective bath, approximated by a  few orbitals \cite{Krauth}. 
It is to be noted that in the metallic state, according to the Fermi liquid theory,
the increase of the on-site interaction leads to an increase in  the renormalized mass of electrons and, consequently,  they tend to be localized. This is the basis of Brinkman-Rice theory of MIT \cite{Brinkman}. However, this scenario can not be applicable for the massless electrons in Graphene's honeycomb lattice. To overcome this problem,  the renormalized Fermi velocity is instead used which is proportional to quasi-particle weight at low energy \cite{Jafari,Tran}. As a case study, we consider a toy density of states (TDOS) model and  examine the effects of interaction on shape of density of states (DOS) and phase transition point. The bar toy DOS has same properties as DOS of honeycomb lattice at $U=0$. In critical $U \sim 11.5$ we can see a gap in DOS at zero energy, implying MIT.   The other quantity which is obtained is renormalized Fermi velocity. At $U\sim 11.5$,  the massless electrons become localized and renormalized Fermi velocity goes to zero. We  have further used ED method on honeycomb lattice to analyze MIT in this system. The ground state of impurity model and the occupancy at  impurity site, at $T=0$ are obtained. For $U_c \sim 11.5$, it is shown that  MIT takes place.\\ 
\section{Model and Method}
The two dimensional Hubbard model on the honeycomb  lattice (HCL) is given by
\begin{equation}
H=-t\sum_{<i,j>,\sigma}c_{i\sigma}^{\dag}c_{j\sigma} -\mu \sum_{i\sigma}c_{i\sigma}^{\dag}c_{i\sigma}
+U\sum_{i}n_{i\uparrow}n_{i\downarrow}
\label{HH}
\end{equation}
where $t$ denotes the nearest neighbor hopping amplitude and  $c_{i\sigma}^{\dag}(c_{i\sigma})$
are the creation (annihilation) operator of electrons at site i with spin $\sigma$. $n_{i,\sigma}$
is the number operator and determines the number of electrons at site i with projection of spin 
$\sigma$. $\mu$ is the chemical potential and $U$ designates the on-site electron-electron repulsion
energy. We take $t$ as the energy unit through out this paper,  we will consider the half filling 
case ($\mu$ = $U$/2).

As already mentioned  DMFT \cite{Georges-RMP} is exact in the limit of infinite lattice coordinations, where 
nonlocal correlations become frozen. In finite dimension this correlations exist but are neglected by DMFT method.
The use of DMFT for study of 2D Hubbard model on honeycomb lattice may accordingly be questionable since this lattice
has only 3 coordinations. However, CDMFT applied on a square lattice shows
that the MIT in 2D lattice are captured by single site DMFT \cite{Wu,CDMFT-2D}. In limit of infinite dimensions, the self-energy thus becomes local function and independent of the momentum but in finite dimension
it can be neglected. The honeycomb lattice is a bipartite
system so the self-energy is written in DMFT as
\begin{eqnarray}
\Sigma_{loc}(\vk, i\omega_{n}) \sim \Sigma(i\omega_{n}),~~
\Sigma_{nloc}(\vk, i\omega_{n}) \sim 0,
\end{eqnarray}
where $\omega_{n}=\frac{(2n+1)\pi}{\beta}$ is Matsubara frequency and the interacting single site Green's function has the following matrix form,
\begin{equation}
G^{-1}(\vk,i\omega_n) =
\left( \begin{array}{cc}
i\omega_n +\mu-\Sigma(i\omega_{n}) & -\epsilon(\vk) \\
-\epsilon^{*}(\vk) & i\omega_n +\mu-\Sigma(i\omega_{n})
\end{array} \right) ,
\label{egf}
\end{equation} 
where $G_{0}^{-1}(\vk,i\omega_n)$ is the non interacting Green's function \cite{Tran}
\begin{equation}
G_{0}^{-1}(\vk,i\omega_n) =
\left( \begin{array}{cc}
i\omega_n +\mu & -\epsilon(\vk) \\
-\epsilon^{*}(\vk) & i\omega_n+\mu
\end{array} \right) 
\label{begf}
\end{equation}
The local Green's function of the original lattice on Hubbard model is obtained by summation over $\vk$
\begin{equation}
G(i\omega_n)=\frac{1}{N}\sum_{\vk}G(\vk,i\omega),
\label{integf}
\end{equation}
and  we can further convert summation to integration $\frac{1}{N}\sum_{\vk} \rightarrow \int{\rho(\epsilon)d\epsilon}$, where
$\rho(\epsilon)$ is the bare density of state \cite{Krauth}.
In DMFT formalism one can  describe 2D Hubbard model by the Anderson impurity model
to describe the dynamics of the local impurity on site \textit{i}  coupled to free conduction
electrons in the bath. The Anderson impurity Hamiltonian is
\begin{equation}
H_{\rm{AIM}} =\sum_{l\sigma} \varepsilon_l c^{\dagger}_{l\sigma} c_{l\sigma} +\sum_{l\sigma} V_{l} (c^{\dagger}_{l\sigma} c_{i\sigma} + c_{l\sigma} c^{\dag}_{i\sigma} ) -\mu c^{\dagger}_{i\sigma} c_{i\sigma} + U n_{i\uparrow} n_{i\downarrow}  ,
\label{aim}
\end{equation}
where $l$ is the  index for bath levels and $i$ corresponds to  impurity level. The impurity Green's function $G^{imp}(i\omega_n)$,  is calculated by the exact diagonalization of the Anderson Hamiltonian. We used the Dyson's equation for calculation of the bare Green's function $G_{0}(i\omega_n)$ which represents effective mean field acting on impurity site
\begin{equation}
G(i\omega_n)=\frac{1}{G^{-1}_{0}(i\omega_n)-\Sigma(i\omega_n)}.
\label{DesonEq}
\end{equation}
In ED method at $U=0$, the impurity Green's function is approximated by the finite numbers of bath levels
\begin{eqnarray}
G^{imp}_{0}(i\omega_n)^{-1} = i\omega_n + \mu - \sum_{l=1}^{n_s} \frac{|V_l|^2}{i\omega_n - \varepsilon_l}, 
\label{impbg}
\end{eqnarray}
where $n_s$ is the bath conduction levels, in the present study we consider $n_s=8$. The Anderson parameters set $\lbrace\varepsilon_{l}, V_{l}\rbrace$ is chosen
such that the \textit{distance function} between the continuous bath function (original lattice)
$G_{0}(i\omega_n)$ and the discretized bath function (impurity model) $G^{imp}_{0}(i\omega_n)$ is minimized.
\begin{eqnarray}
d=\frac{1}{n_{max}+1} \sum_{n=0}^{n_{max}} |\omega_n|^{-k}|G^{imp}_{0}(i\omega_n)^{-1} - G_{0}(i\omega_n)^{-1}|^2.
\label{disf}
\end{eqnarray}
where $n_{max}$ is a large upper cutoff, (here $n_{max}=2^{12}$).The parameter $k$  is very
important when we have small number of bath levels, and if chosen large (e.g. k = 3) ,
enhances the importance of the lowest Matsubara frequencies in the
minimization procedure \cite{Tran,Katsnelson}. Above equations are the basic equations for the the self-consistent
solution in ED method. Below we describe our approach in more detail. \\

 First we use initial set of Anderson parameters, then $G^{imp}_{0}(i\omega_n)$ and $G^{imp}(i\omega_n)$ are calculated
 by (\ref{impbg}) and diagonalization by (\ref{aim}), respectively. The self-energy is obtained by Dyson equation, ($\Sigma=G^{imp}_{0}-G^{imp}$). In second step $G(i\omega_n)$ is calculated by (\ref{egf}-\ref{integf}), new bar 
 Green's function is constructed from (\ref{DesonEq}).  For the sake of faster convergence one could use $G^{new}_{0}=\alpha G_{0}+(1-\alpha)G^{imp}_{0}$ ($0\leq \alpha \leq 1$). At  final step, a new Anderson parameters  is obtained by minimization
 of distance function \ref{disf} ($G^{new}_{0}(i\omega_n) \rightarrow G_{0}(i\omega_n)$). This cycle is repeated until 
 $G_{0}(i\omega_n) \sim G^{imp}_{0}(i\omega_n)$ and $G(i\omega_n) \sim G^{imp}(i\omega_n)$. After convergence, it is possible to 
 extract the desired observable and physical quantities \textit{e.g.} DOS,
 quasi-particle weight and double occupancy.

\section{Results and Discussions}
\begin{figure}[tb]
  \begin{center}
    \includegraphics[width=9cm,angle=0]{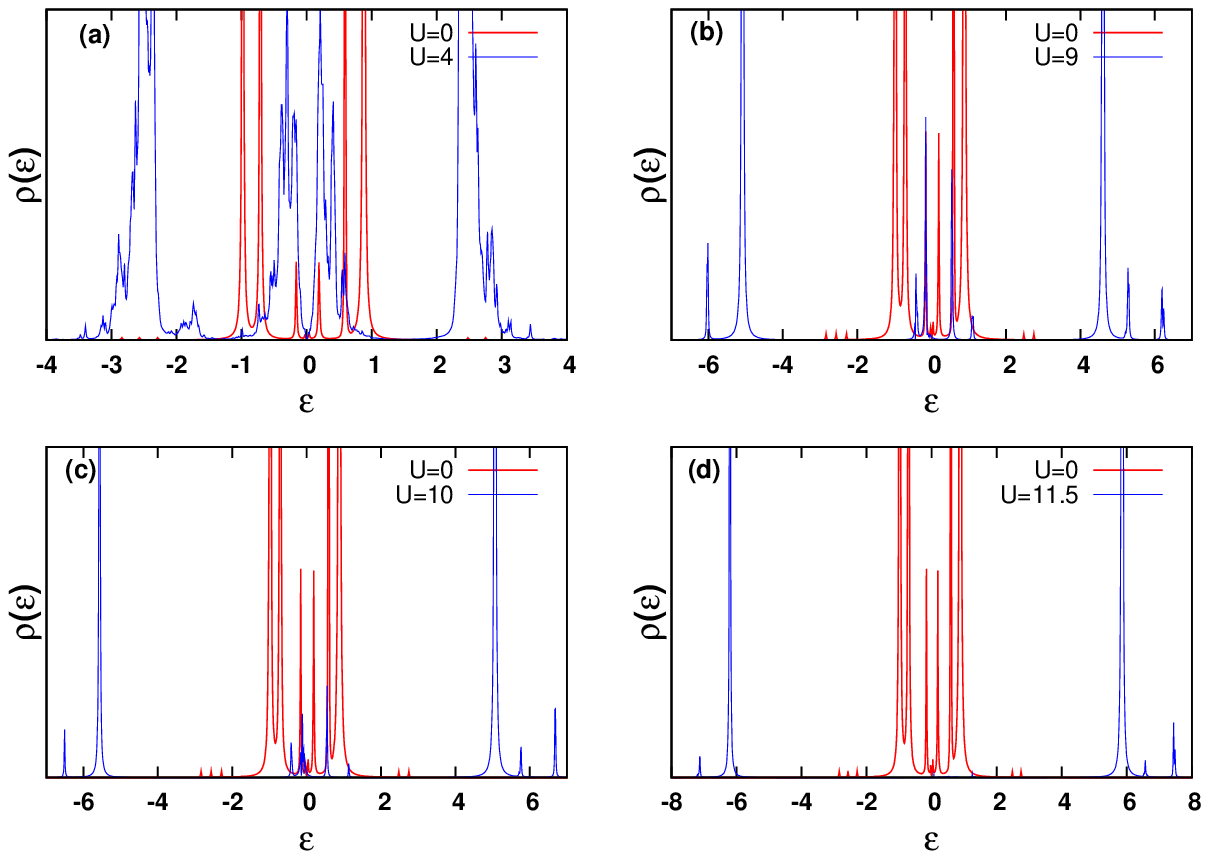}
    \caption{(Color online) The comparison between non interacting (red)
    and interacting(blue) DOS in 2D Hubbard on toy model. By increasing 
    on site interaction singularity  moves toward low energy, and two Hubbard bands 
    appear in two sides of low energy.
    } 
    \label{TDOS.fig}
  \end{center}
\end{figure}

In this paper we first calculated the local Green's function of the honeycomb lattice (\ref{integf}).
The density of states per unit cell for non interacting honeycomb lattice is
\begin{eqnarray}
   \rho(\epsilon)&=&\frac{|\epsilon|}{\pi^2} \frac{1}{\sqrt{Z_0}}
   F\left(\frac{\pi}{2},\sqrt{\frac{Z_1}{Z_0}} \right),\nn\\
   Z_0&=&\left\{
   \begin{array}{lr}
      \left(1+|\epsilon|\right)^2-\left(\epsilon^2-1\right)^2/4, & |\epsilon|\le1\\
      4|\epsilon|,					& 1\le |\epsilon|\le 3
   \end{array}
   \right.\nn\\
  Z_1&=&\left\{
   \begin{array}{lr}
      4|\epsilon|,	& |\epsilon|\le1\\
      \left(1+|\epsilon|\right)^2-\left(\epsilon^2-1\right)^2/4, & 1\le |\epsilon|\le 3\\
      \end{array}
   \right.,
   \label{DOScom}
\end{eqnarray}
where $F(\pi/2,x)$ is the complete elliptic integral of the first kind. Close to the Dirac point,
the dispersion is approximated by $\epsilon(k)=\pm v_{F}k$ and DOS is
\begin{equation}
\rho(\varepsilon)=\frac{2A_c}{\pi}\frac{\left|\epsilon \right|}{v^{2}_{F}}
\label{DOSapp}
\end{equation}
where $A_c$ is the unit cell area, and $v_{F} = 3ta/2$ is the Fermi velocity \cite{NetoRMP}. In the following  a toy DOS (which is conceptually similar to DOS of Graphene and basically represents the same properties \cite{Jafari})  is used,
\begin{equation}
   \rho(\varepsilon)=
   \left\{ \begin{array}{ll}
   \frac{2A_c}{\pi}\frac{\left|\epsilon \right|}{v^{2}_{F}} & |\epsilon|\le3,\ne1 \\
   1              & |\epsilon|=1\\
   0                 & |\epsilon| > 3
   \end{array}\right..
   \label{dosmodel.eqn}
\end{equation}
In ED method DOS is approximated by a set of delta functions. This is because in this method
a finite number of conduction levels in the effective bath are used. The calculated results are accordingly compared with those reported in Ref. \cite{Jafari}.
Insert TDOS in DMFT process and increasing on-site interaction, the
upper and lower Hubbard bands are made and the singularity (for similarity
with Van-Hove singularity) is moved to lower energy, see Fig.\ref{TDOS.fig}.  As expected, because of 
delta functions shape of DOS, we could not see 
any change in Fermi velocity by increasing $U$ \cite{Jafari,Tran}.

\begin{figure}[tb]
  \begin{center}
   \includegraphics[width=9cm,angle=0]{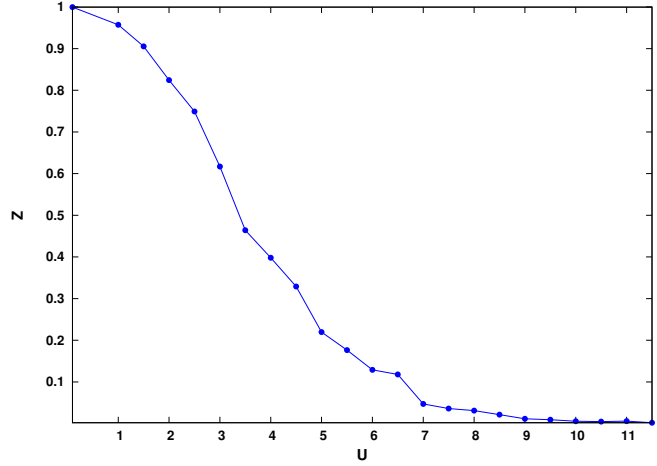}
    \caption{(Color online) The renormalized Fermi velocity decrease
    by increasing $U$, near $U_{c}\sim 11$ Fermi velocity becomes zero and
    semimetal insulator transition (SMIT) occur.
    } 
    \label{Zf.fig}
  \end{center}
\end{figure}

\begin{figure}[tb]
  \begin{center}
    \includegraphics[width=9cm,angle=0]{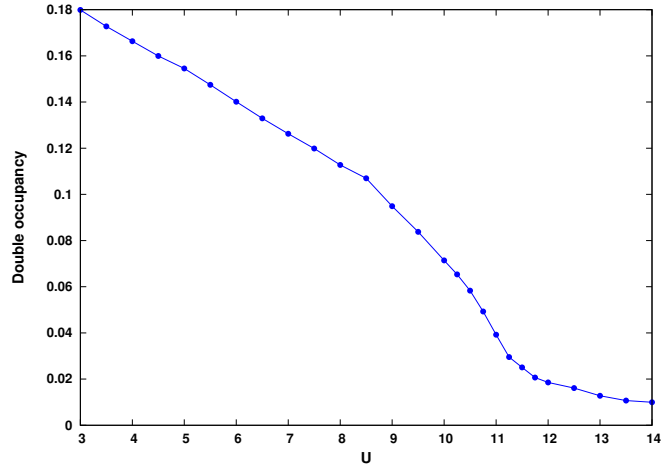}
    \caption{(Color online) The double occupancy of Honeycomb lattice
    for various $U$. As shown the curvature of $D$  changes near $U_c\sim 11\rightarrow 11.5$ indicating 
    a second order phase transition.
    } 
    \label{docg.fig}
  \end{center}
\end{figure}

\begin{figure}[tb]
  \begin{center}
    \includegraphics[width=9cm,angle=0]{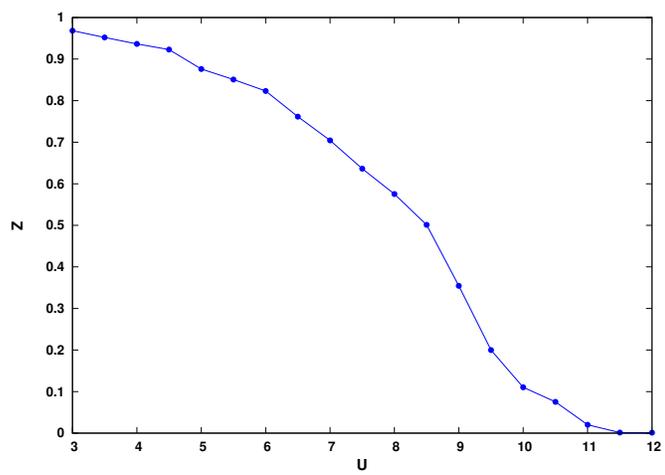}
    \caption{(Color online) The Fermi velocity renormalized factor for Honeycomb lattice near $U_{c} \sim 11.5$. 
    } 
    \label{zg.fig}
  \end{center}
\end{figure}

The quasi-particle weight is the physical quantity for analysis of MIT according to
Brinkman-Rice theory \cite{Brinkman}, But in the honeycomb lattice quasi-particles are massless (we have SMIT),
so this theory could not be applicable here,  instead we have used the renormalized Fermi velocity factor \cite{Jafari,Tran}
which have the same role in SMIT, 
\begin{equation}
Z=\frac{\displaystyle 1}{\displaystyle 1-\frac{\partial{\Re}\Sigma(\omega+i\eta)}{\partial \omega}
\bigg|_{\omega=0}}.
\label{Zf}
\end{equation}
Since the slope of $\Im \Sigma(i\omega)$ and $\Re \Sigma(\omega+i\eta)$ are identical in $\omega \rightarrow 0$
limit, so the ${\Im \Sigma(i\omega_n)}/{\omega_n}|_{n=1}$ and
${\partial \Re(\omega +i\eta)}/{\partial \omega}|_{\omega=0}$ have similar treatments at low frequency.
In Fig. \ref{Zf.fig} we plot Fermi velocity renormalized of TDOS for weak and strong couplings.  It is  found 
that for $U_c \sim 11 \rightarrow 11.5$ this factor goes to zero.  This shows that the system is in the insulating phase \cite{Tran,Jafari}.\\
   The final result of ED for toy model predicts $U_c \sim 11-11.5$ for SMIT but in previous DMFT calculations
by IPT method $U_c$ is found to be  larger. This is due to the fact that  IPT usually overestimates this critical value \cite{Vollhardt}.
We next compute  the Green's function  of system $G(\omega+i\eta)$
using continued-fraction expansion and ground-state by Lanczos method. The double occupancy is then obtained directly by $D=\langle g.s.|n_{\uparrow}n_{\downarrow}|g.s.\rangle$ \cite{Georges-RMP}.
The  $\partial D/\partial U$ is proportional to susceptibility, $\chi$, 
so when curvature of $D$ changes,  a singularity occurs in susceptibility of the system, implying that the system
has undergone a phase transition.  We already know $\chi=\partial^2 F/\partial^2 U$, where $F$ is the free energy. This means that the singularity in
susceptibility corresponds to a second order phase transition  \cite{Sorella}. In Fig.\ref{docg.fig} we have accordingly plotted double occupancy for weak and stronge interactions. The  phase transition point cab be seen near $U_c \sim 11.25$.  
The renormalized Fermi velocity for honeycomb lattice is shown in Fig.\ref{zg.fig}, the critical 
on-site interaction is predicted to be between $\sim 11-11.5$, in accordance with that described for Fig.\ref{docg.fig}.

The previous works based on Hubbard model have found phase transition in the honeycomb lattice at  various $U_c$ values:
The quantum Monte Carlo method used by \cite{Sorella} predicted a semi-metal (SM) to anti ferromagnetic
transition (AFT), followed by a subsequent transition to  Mott insulator at $U_c \sim 4.5 \pm 0.5$.
In the other QMC study considering the broken symmetry through second-order 
perturbation theory, the authors   found the same SM-AFT phase transition   at $U_c=2.3$
for infinite dimensional diamond lattice \cite{Santoro}. The analysis of SMIT in the HCL has also been done by IPT. The respective  SMIT in HCL has been shown to be of  second order but at higher $U_c=13.3$. As mentioned above IPT is known to overestimate
the value of $U_c$  as compared with the other DMFT methods. It is worth mentioning that recently 
Tran and Kuroki   \cite{Tran} predicted both the first and second order phase transition points, and their coexistence in MIT.

\section{Conclusion}
We studied 2D Hubbard model on Honeycomb lattice and analyzed effects of short range
interaction on renormalization of Fermi velocity and phase transition  by DMFT.
We used exact diagonalization method for solving 2D Hubbard model and calculated double occupancy and
renormalized Fermi velocity. The  phase transition point was accordingly obtained.  Using two different schemes we predicted
the same critical on-site interaction $U_c \sim 11.25$.  The phase transition was found to be of the 
second order with a critical $U$ lower than that given in Ref.  \cite{Jafari} but close to that given in Ref.
\cite{Santoro, Tran}.

\newpage
\section*{Figure captions:}

\end{document}